\documentclass[aps,reprint,longbibliography]{revtex4-1}

\usepackage{amsmath}
\usepackage{graphicx}
\graphicspath{{figures/}}
\usepackage{siunitx}

\renewcommand{\vec}[1]{\mathbf{#1}}
\newcommand{\abs}[1]{\left\vert #1 \right\vert}

\begin{document}

\title{Acoustokinetics: Crafting force landscapes from sound waves}

\author{Mohammed A. Abdelaziz} \author{David G. Grier}
\affiliation{Department of Physics and Center for Soft Matter
  Research, New York University, New York, NY 10003}

\date{}

\begin{abstract}
  Factoring the pressure field of a harmonic sound wave into its
  amplitude and phase profiles provides the foundation for an
  analytical framework for studying acoustic forces that not only
  provides novel insights into the forces exerted by specified sound
  waves, but also addresses the inverse problem of designing sound
  waves to implement desired force landscapes.  We illustrate the
  benefits of this acoustokinetic framework through case studies of
  purely nonconservative force fields, standing waves, pseudo-standing
  waves, and tractor beams.
\end{abstract}

\maketitle

\section{Introduction}
\label{sec:introduction}

Structured sound waves exert forces and torques that can be harnessed
to transport, sort and organize insonated objects
\cite{zhang2011angular,wang2012autonomous,marzo2015holographic,marzo2019holographic}.
Applications include non-contact processing of sensitive
\cite{xie2006acoustic,cao2012rapid} and hazardous
\cite{andrade2016acoustic} materials, flow focusing for materials
analysis and medical diagnostics \cite{lenshof2012acoustofluidics},
and automated remote manipulation for research
\cite{courtney2013dexterous}.  Rapidly growing interest in harnessing
acoustic forces has inspired a fundamental reassessment of the physics
of wave-mediated forces.  Recent developments in the theory of
acoustic forces
\cite{bjerknes1906fields,blake1949bjerknes,gorkov62,settnes2012forces,silva14}
parallel the analogous theory of optical forces
\cite{gordon73,chaumet00}.  Both offer valuable and often surprising
insights into the elementary mechanisms of wave-matter interactions.
The acoustokinetic framework introduced here addresses the
complementary inverse problem: identifying what wave will create a
desired force landscape.

The inverse problem for optical forces recently has been rendered more
tractable by expressing the electromagnetic field in terms of its
real-valued amplitudes and phases along each Cartesian coordinate
\cite{ruffner13,yevick16}.  This approach is called the theory of
photokinetic effects and yields useful analytic expressions for the
performance of optical traps \cite{yevick17} including design criteria
for optical tractor beams \cite{yevick16}.  Here, we show that an
analogous factorization of the pressure field in sound waves is
similarly useful for understanding and implementing acoustic
manipulation.  We illustrate the utility of this acoustokinetic
framework through case studies on nonconservative acoustic force
fields, standing and pseudo-standing waves, and acoustic tractor
beams.

\subsection{Light: Photokinetic analysis}
\label{sec:light}

We develop acoustokinetics by analogy to photokinetics and therefore
briefly review the theory of optical forces.  A small particle
immersed in an electromagnetic wave develops an electric dipole moment
proportional to the local field.  This induced dipole experiences a
time-averaged Lorentz force in gradients of the field that can be
expressed as \cite{chaumet00}
\begin{equation}
  \label{eq:chaumet}
  \vec{F}_e(\vec{r})
  =
  \frac{1}{2}
  \Re \left\{
    \alpha_e \sum_{j=1}^3 E_j(\vec{r})
    \nabla 
    E_j^\ast(\vec{r})
  \right\},
\end{equation}
where $E_j(\vec{r})$ is the $j$-th Cartesian coordinate of the
electric field and $\alpha_e$ is the particle's complex dipole
polarizability.  Expressing the components of the electric field in
terms of their real-valued amplitude and phase profiles,
\begin{equation}
  E_j(\vec{r})
  = 
  u_j(\vec{r}) \, e^{i \varphi_j(\vec{r})},
\end{equation}
yields the surprisingly simple expression \cite{ruffner13},
\begin{equation}
  \vec{F}_e(\vec{r})
  =
  \frac{1}{4} \alpha_e' 
  \nabla \sum_{j = 1}^3 u_j^2(\vec{r})
  +
  \frac{1}{2} \alpha_e''
  \sum_{j = 1}^3
  u_j^2(\vec{r}) 
  \nabla \varphi_j(\vec{r}),
  \label{eq:photokinetic}
\end{equation}
where $\alpha_e'$ and $\alpha_e''$ are the real and imaginary parts of
the polarizability, respectively.

The first term on the right-hand side of Eq.~\eqref{eq:photokinetic}
is the manifestly conservative intensity-gradient force responsible
for single-beam optical traps such as optical tweezers
\cite{ashkin86}.  The second describes a nonconservative force
\cite{wu09} that is directed by phase gradients \cite{roichman08}.
Phase-gradient forces tend to drive trapped particles out of
thermodynamic equilibrium with their supporting media
\cite{roichman08,sun09}, mediate the transfer of light's orbital
angular momentum \cite{allen92,he95a,gahagan96,simpson96}, and have
been used to create light-driven micromachines such as pumps
\cite{ladavac04}, mixers \cite{ladavac05} and optical tractor beams
\cite{lee10}.  Even non-absorbing dielectric particles experience
nonconservative optical forces because of radiative contributions to
the dipole polarizability \cite{draine88,albaladejo10}.

The dipole-order expression in Eq.~\eqref{eq:photokinetic} accurately
describes the forces experienced by particles with radii, $a_p$, that
are small enough to satisfy the Rayleigh criterion, $k a_p < 1$, where
$k$ is the wavenumber of light.  In the Rayleigh regime, the
conservative intensity-gradient force generally dominates the
light-matter interaction because $\alpha_e'$ scales as $(k a_p)^3$,
whereas $\alpha_e''$ scales as $(k a_p)^6$.

\subsection{Sound: Acoustic radiation forces}
\label{sec:sound}

The analogous dipole-order acoustic radiation force experienced by a
small particle in a harmonic sound field may be expressed in terms of
the pressure, $p(\vec{r}, t)$, as \cite{silva14}
\begin{equation}
  \label{eq:silva}    
  \vec{F}(\vec{r}) 
  =
  \frac{1}{2} \Re\left\{ 
    \alpha_a p \nabla p^\ast
    +
    \beta_a k^{-2} (\nabla p \cdot \nabla) \nabla p^\ast
  \right\} ,
\end{equation}
where the coefficients $\alpha_a$ and $\beta_a$ play the role of
dipole and quadrupole polarizabilities, respectively.  This expression
is analogous to Eq.~\eqref{eq:chaumet} for optical forces and is
obtained by rearranging terms from Eq.~(16) in Ref.~\cite{silva14}.
It therefore also is equivalent \cite{silva14} to the angular spectrum
decomposition of $\vec{F}(\vec{r})$ \cite{sapozhnikov2013radiation}
for $k a_p < 1$.  Expressing $\vec{F}(\vec{r})$ in terms of multipole
polarizabilities clarifies the analogy with photokinetics.  Lengths in
Eq.~\eqref{eq:silva} are scaled by the wavenumber, $k = \omega/c_m$,
where $\omega$ is the sound's frequency and $c_m$ is its speed in the
medium.  Equation~\eqref{eq:silva} applies to inviscid fluids, for
which the pressure satisfies the scalar wave equation
\begin{equation}
  \label{eq:waveequation}
  \nabla^2 p = -k^2 p.
\end{equation}

Our focus on traveling waves in inviscid media is inspired by our
interest in developing new modalities of long-ranged non-contact
manipulation.  Long-range manipulation is facilitated by minimizing
acoustic losses in the medium.  This can be achieved in air by working
at frequencies below \SI{50}{\kilo\hertz}
\cite{marzo2015holographic,lim2019edges}, for which the acoustic
attenuation is less than \SI{2}{\deci\bel\per\meter} under standard
conditions \cite{bass1995atmospheric} and scales as $\omega^2$ for
lower frequencies.  The equivalent limiting frequency for water is
roughly \SI{2}{\mega\hertz} \cite{ainslie1998simplified}.  Working at
low frequencies also minimizes the influence of acoustic streaming
forces, which ordinarily compete with acoustic radiation forces in
viscous media and in inviscid media bounded by confining surfaces
\cite{lighthill1978acoustic}.

An object's dipole and quadrupole polarizabilities generally depend on
its size, shape and composition as well as the frequency of the sound
and the properties of the fluid medium.  For simplicity and
concreteness, we will specialize to the case of a spherical scatterer
of radius $a_p$ that is composed of a material of density $\rho_p$ and
sound speed $c_p$.  Such an object's response to the sound field is
characterized by the polarizabilities \cite{silva14}
\begin{subequations}
  \label{eq:alphabeta}
  \begin{align}
    \label{eq:alpha}
    \alpha_a
    & =
      \frac{4\pi a_p^3}{3\rho_m c_m^2} \, f_0 \,
      \left[
      -1 + i \frac{1}{3} (f_0 + f_1) (k a_p)^3 
      \right] \\
    \label{eq:beta}
    \beta_a 
    & = 
      \frac{2\pi a_p^3}{\rho_m c_m^2} \, f_1 \,
      \left[1 + i \frac{1}{6} f_1 (k a_p)^3 \right],
  \end{align}
\end{subequations}
where the monopole coupling coefficient,
\begin{subequations}
  \label{eq:couplingcoefficients}
  \begin{equation}
    \label{eq:compressibility}
    f_0 = 1 - \frac{\rho_m c_m^2}{\rho_p c_p^2},
  \end{equation}
  depends on the compressibility mismatch between the particle and the
  medium, and the dipole coupling coefficient,
  \begin{equation}
    \label{eq:densitymismatch}
    f_1 = 2\frac{\rho_p - \rho_m}{2 \rho_p + \rho_m},
  \end{equation}
\end{subequations}
gauges the density mismatch.  These expressions also are obtained by
reorganizing coefficients from Eq.~(16) of Ref.~\cite{silva14} and are
valid for $k a_p < 1$.  They constitute the leading-order
contributions for both the real parts of the polarizabilities,
$\alpha_a'$ and $\beta_a'$, and also the imaginary parts, $\alpha_a''$
and $\beta_a''$.

\begin{figure}
  \centering \includegraphics[width=0.8\columnwidth]{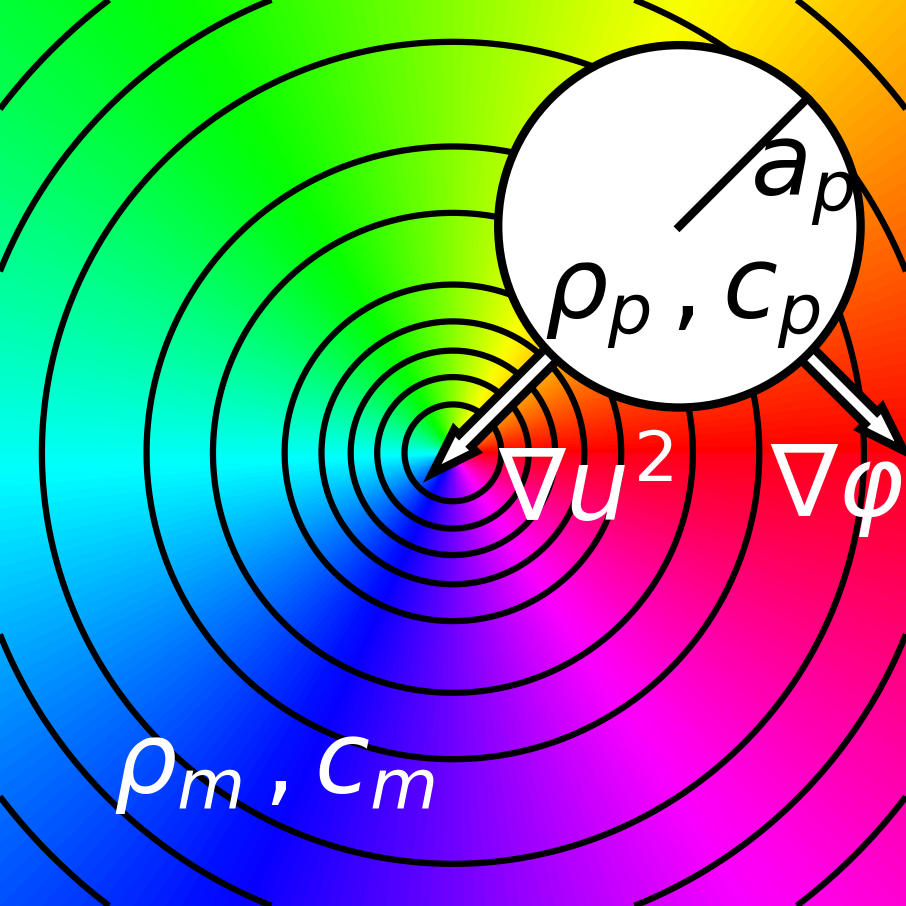}
  \caption{Schematic representation of a sphere of radius $a_p$
    immersed in an acoustic pressure field.  Contours denote
    isosurfaces of the pressure intensity. Colors represent the phase
    of the pressure field.  Generally speaking, intensity gradients
    direct conservative trapping forces while phase gradients direct
    nonconservative driving forces.}
  \label{fig:acoustokinetic}
\end{figure}

\section{Acoustokinetic Framework}
\label{sec:acoustokineticappendix}

Drawing on the analogy with photokinetics, we express the harmonic
sound wave's pressure field in terms of its amplitude and phase
profiles:
\begin{equation}
  \label{eq:pressureprofile}
  p(\vec{r}, t) = u(\vec{r}) \, e^{i \varphi(\vec{r})} \, e^{-i \omega t}.
\end{equation}
The first term on the right-hand side of Eq.~\eqref{eq:silva} then
yields
\begin{subequations}
  \label{eq:acoustokinetic}
  \begin{equation}
    \label{eq:falpha}
    \vec{F}_\alpha(\vec{r})
    =
    \frac{1}{4} \alpha_a' \, \nabla u^2
    +
    \frac{1}{2} \alpha_a'' \, u^2 \nabla \varphi,
  \end{equation}
  which is directly analogous to Eq.~\eqref{eq:photokinetic} for the
  dipole-order force exerted by light.  These contributions to the
  acoustic radiation force are depicted in
  Fig.~\ref{fig:acoustokinetic}.  As in the optical case, $\alpha_a'$
  and $\alpha_a''$ scale as $(ka_p)^3$ and $(ka_p)^6$ respectively,
  which means that the conservative force generally dominates for
  small particles.

  The second term on the right-hand side of Eq.~\eqref{eq:silva}
  arises from the velocity-matching condition at the sphere's boundary
  and so has no analogue in optical radiation forces.  It vanishes for
  density-matched particles ($\beta_\alpha = 0$), which therefore
  behave exactly like dielectric particles in a light field, in
  agreement with conclusions from previous studies
  \cite{toftul2019acoustic}.  When expressed in terms of the amplitude
  and phase profiles, this term separates naturally into a
  conservative contribution,
  \begin{equation}
    \label{eq:fbetaconservative}
    \vec{F}_\beta^c(\vec{r})
    =
    \frac{1}{4} \beta_a'
    \nabla \left( u^2 + \frac{1}{2} k^{-2} \nabla^2 u^2 \right) ,
  \end{equation}
  that augments the conservative intensity-gradient force from
  $\vec{F}_\alpha(\vec{r})$ and a nonconservative contribution,
  \begin{multline}
    \label{eq:fbetanonconservative}
    \vec{F}_\beta^{nc}(\vec{r}) = \frac{1}{4} \beta_a'' k^{-2} \left[
      (2k^2u^2+\nabla^2 u^2 + 2 u \nabla u \cdot \nabla) \nabla
      \varphi \right. \\ \left.  - (u \nabla^2 \varphi + 2 u \nabla
      \varphi \cdot \nabla) \nabla u \right],
  \end{multline}
  that is directed both by phase gradients and also by amplitude
  gradients.  The combination,
  \begin{equation}
    \vec{F}_\beta(\vec{r})
    =
    \vec{F}_\beta^c(\vec{r})
    +
    \vec{F}_\beta^{nc}(\vec{r}),
  \end{equation}
\end{subequations}
captures the sphere's leading-order coupling to the quadrupole
components of the incident field.  A derivation of
Eq.~\eqref{eq:acoustokinetic} from Eq.~\eqref{eq:silva} is presented
in Appendix~\ref{sec:appendix}.

Unlike the optical case, where quadrupolar forces generally are weaker
than dipole contributions, the two terms in $\vec{F}_\beta(\vec{r})$
can be comparable in magnitude to their counterparts in
$\vec{F}_\alpha(\vec{r})$ because $\beta_a'$ scales as $(k a_p)^3$ and
$\beta_a''$ scales as $(k a_p)^6$.  These density-dependent terms
therefore can be used to exert control in ways that are not possible
with light.

For very small particles satisfying $ka_p \ll 1$, the acoustic force
field is dominated by the conservative terms proportional to
$\alpha_a'$ and $\beta_a'$.  These terms are identical to the force
described by the classic Gor'kov potential \cite{gorkov62}, which is
widely used to describe acoustic trapping phenomena
\cite{marzo2015holographic,bazou2012controlled}.  For larger
particles, and for appropriately structured sound fields,
non-conservative contributions proportional to $\alpha_a''$ and
$\beta_a''$ can be significant, and even can be dominant
\cite{melde2016holograms,demore14}.  Such contributions are not
accounted for by the Gor'kov potential.

The acoustokinetic framework described by
Eq.~\eqref{eq:acoustokinetic} is the principal contribution of this
work.  We now demonstrate its value through case studies on realizable
sound fields with exceptional properties.

\section{Applications of the Acoustokinetic Framework}
\label{sec:applications}

\subsection{Designing purely nonconservative force fields}
\label{sec:nonconservative}

To illustrate how the acoustokinetic framework can address the inverse
problem of designing sound waves to implement desired force
landscapes, we use Eq.~\eqref{eq:acoustokinetic} to design harmonic
sound waves that exert purely nonconservative forces.  This is
equivalent to requiring the conservative part of the acoustic
radiation force to vanish, and thus requires us to look beyond the
Gor'kov potential.  Equations~\eqref{eq:falpha} and
\eqref{eq:fbetaconservative} show that this goal can be met if the
particle is not density matched, $\beta_a' \neq 0$, and if the
pressure intensity satisfies the inhomogeneous Helmholtz equation,
\begin{equation}
  \label{eq:helmholtz}
  \nabla^2 u^2 + 2 \left(1 + \frac{\alpha_a'}{\beta_a'} \right) k^2 \, u^2
  = C .
\end{equation}
The undetermined constant $C$ distinguishes families of
non-conservative sound waves for the class of objects with compatible
values of $\alpha_a'/\beta_a'$.  Solutions to Eq.~\eqref{eq:helmholtz}
must be real-valued and must be paired with real-valued phase profiles
that complete the description of the pressure field and satisfy the
wave equation, Eq.~\eqref{eq:waveequation}.

\begin{figure}
  \centering \includegraphics[width=0.8\columnwidth]{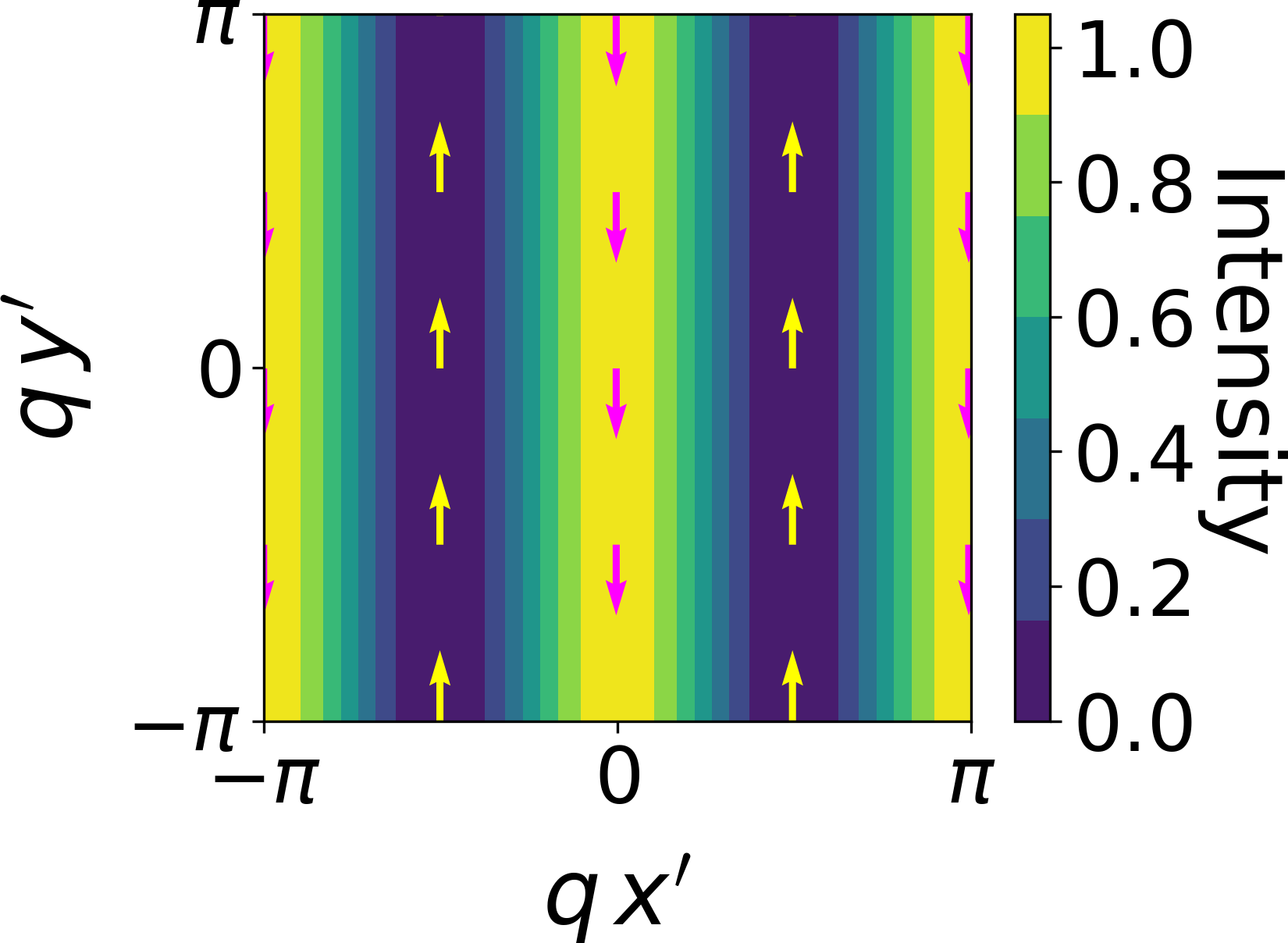}
  \caption{Intensity of the ``picket fence'' field in the
    $x'y'$-plane. For particularly selected parameters, the
    conservative force vanishes everywhere in this field, and the
    remaining force (indicated by arrows) is purely
    nonconservative. With the choice of material properties plotted
    ($C_2/C_1=1/3$), the direction of the force is also spatially
    modulated. The sign of the force in the $\hat y'$ direction is
    also its sign in the $\hat z$ direction.}
  \label{fig:nonCon}
\end{figure}

\begin{subequations}
  One interesting set of purely nonconservative solutions has the
  sinusoidal amplitude profile
  \begin{equation}
    \label{eq:nonconservativeamplitude}
    u(\vec{r}) = p_0 \cos(q(x-y)),
  \end{equation}
  with spatial frequency
  \begin{equation}
    \label{eq:nonconservativeq}
    q 
    = 
    \frac{1}{2} 
    \sqrt{1 + \frac{\alpha_a'}{\beta_a'}} \, k.
  \end{equation}
  The associated phase profile,
  \begin{equation}
    \label{eq:nonconservativephase} 
    \varphi(\vec{r})
    =
    kz \, \cos\gamma + (x + y)
    \sqrt{ \frac{1}{2} k^2 \sin^2\gamma - q^2},
  \end{equation}
  identifies this field as the superposition of two plane waves, each
  oriented at angle $\gamma$ relative to $\hat{z}$ and at angle
  \begin{equation}
    \theta = 
    \cos^{-1}\left(-\frac{\alpha_a'}{\beta_a'}\right)
  \end{equation}
  relative to one another in the $(x,y)$-plane.
\end{subequations}
Under these conditions, the in-plane component of the radiation
pressure exactly cancels the conservative intensity-gradient force.
The remaining scattering force is sinusoidally modulated in the
transverse plane.  The result is a ``picket fence'' of parallel force
lines, which is illustrated in Fig.~\ref{fig:nonCon} using the rotated
coordinates
\begin{subequations}
  \begin{align}
    x' & = x - y \quad \text{and}\\
    y' & = x + y
  \end{align}
\end{subequations}
for clarity.  In these coordinates, the net force,
\begin{subequations}
  \label{eq:nonconservativeforce}
  \begin{equation}
    \vec{F}(\vec{r}')
    = 
    \frac{1}{4} k p_0^2 \, f(x') \, \hat{F},
  \end{equation} 
  is directed along
  \begin{equation}
    \hat{F}
    =
    \sqrt{\frac{1}{2}\sin^2\gamma - \frac{q^2}{k^2}}
    \, \hat{y}' 
    + \cos\gamma \, \hat{z},
  \end{equation}
  and has an amplitude that varies with the transverse coordinate as
  \begin{equation}
    f(x') 
    =
    C_1 + 2(C_2-C_1)\cos^2(qx').
  \end{equation} 
  The scale and depth of the force landscape's modulation depend on
  the object's properties through
  $C_1 = \beta_a''(1 + \alpha_a'/\beta_a')$ and
  $C_2 = \alpha_a'' + \beta_a''$.
\end{subequations}

Picket fence modes for a given type of object are distinguished by the
angle of inclination, $\gamma$.  The range of possible angles is
limited by Eq.~\eqref{eq:nonconservativephase} to
\begin{equation}
  \sin^2\gamma > \frac{2 q^2}{k^2}.
\end{equation}
With this constraint, picket fences can be created for objects that
satisfy
\begin{equation}
  \label{eq:nonconservativecondition}
  -1 < \frac{\alpha_a'}{\beta_a'} < \sin^2\gamma.
\end{equation}
Under some conditions, including those depicted in
Fig.~\ref{fig:nonCon}, the direction of the force can alternate within
the fringe pattern.  In terms of the standard coupling coefficients,
alternating picket fences can be projected for objects satisfying
\begin{equation}
  -\frac{3}{4} < \frac{f_0}{f_1} < -\frac{1}{2}.
\end{equation}
Droplets of xylene hexafluoride
($\rho_p = \SI{1370}{\kg\per\cubic\meter}$,
$c_p = \SI{880}{\meter\per\second}$) dispersed in butanol
($\rho_m=\SI{810}{\kg\per\cubic\meter}$,
$c_m=\SI{1240}{\meter\per\second}$), for example, have
$f_0/f_1 = \num{-0.55}$ and thus are predicted to experience an
alternating picket fence force field.  Although alternating picket
fences are only possible for a limited domain of material properties,
picket fences in general could have practical applications for sorting
objects by density or compressibility.

\subsection{Conservative forces in standing waves}
\label{sec:standingwaves}

\begin{figure}[t]
  \centering
  \includegraphics[width=0.8\columnwidth]{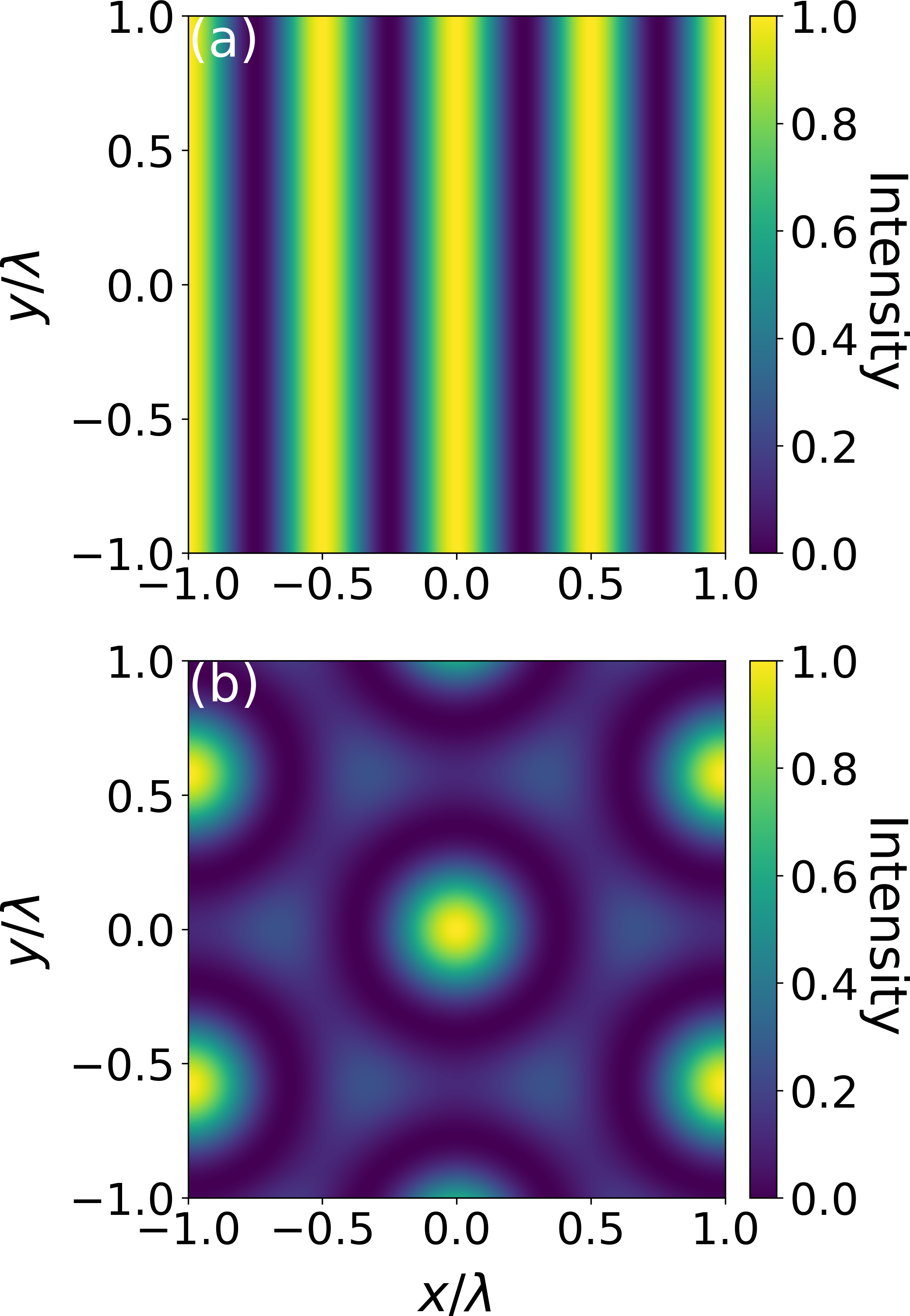}
  \caption{Maps of the pressure intensity of (a) a one-dimensional
    standing wave and (b) a sixfold standing wave.  Particles are
    trapped either at intensity maxima or minima, depending on their
    composition relative to the medium.}
  \label{fig:intplots}
\end{figure}

The acoustokinetic framework also is useful for analyzing the force
fields created by specified sound waves.  Standing waves, for example,
can be decomposed into superpositions of counterpropagating plane
waves.  The phase-dependent terms in $\vec{F}_\alpha(\vec{r})$ and
$\vec{F}_\beta(\vec{r})$ vanish in such superpositions, leaving a
manifestly conservative force landscape,
\begin{equation}
  \label{eq:standingwave}
  \vec{F}_\text{standing}(\vec{r})
  =
  \frac{1}{4} \nabla
  \left[ \alpha_a' u^2 + \beta_a' k^{-2} (\nabla u)^2 
  \right].
\end{equation}
Particles satisfying
\begin{equation}
  \label{eq:standingantinode}
  \alpha_a' - \beta_a' > 0   
\end{equation}
are drawn toward antinodes of the pressure field, as illustrated in
Fig.~\ref{fig:intplots}.  This condition is satisfied by compressible,
low-density particles such as bubbles in water.  Particles with
complementary properties are drawn toward nodes.

\subsection{Nonconservative forces in pseudo-standing waves}
\label{sec:pseudostanding}

\begin{figure}
  \centering \includegraphics[width=\columnwidth]{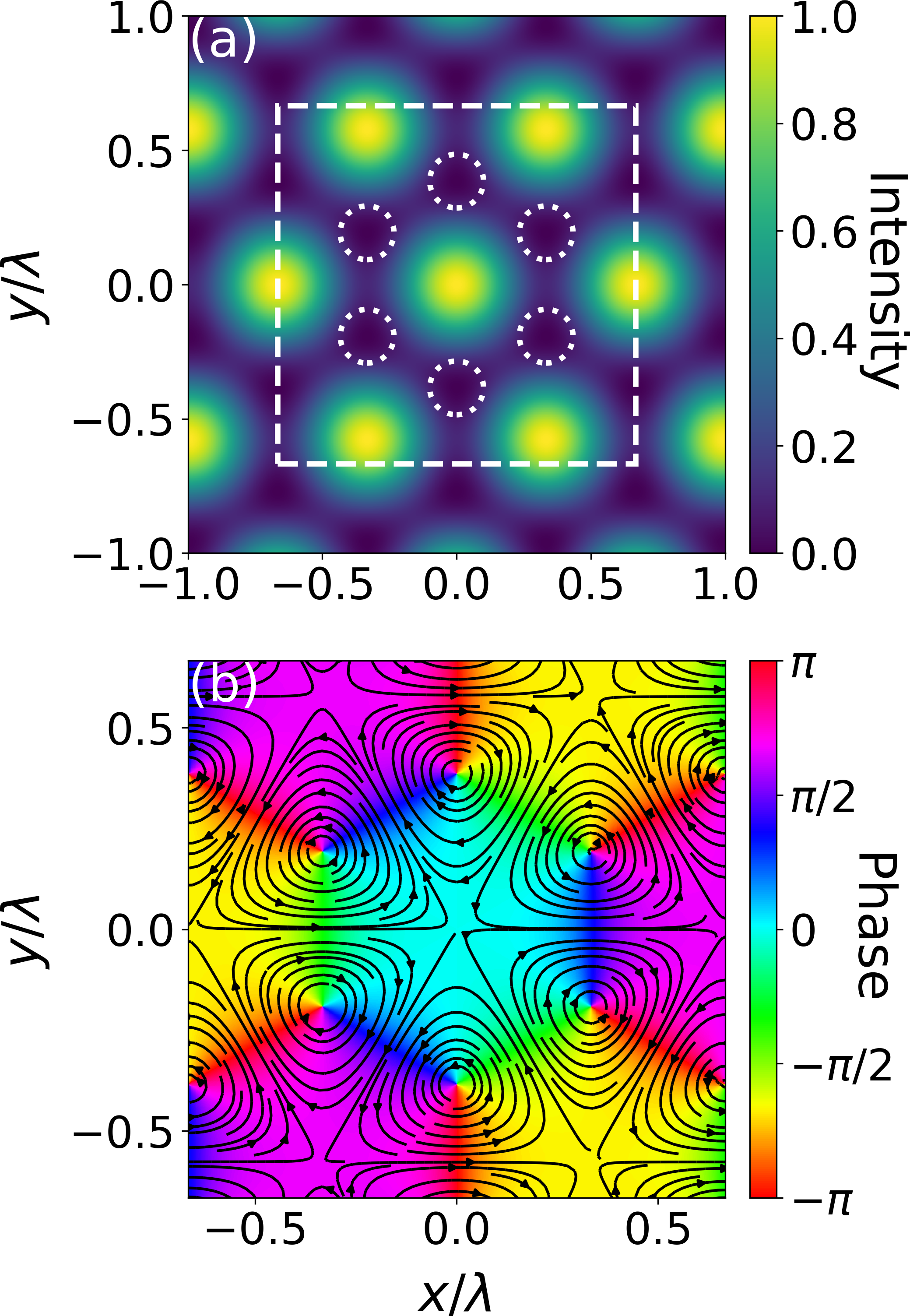}
  \caption{(a) Intensity and (b) phase maps of the threefold
    zero-momentum wave. Dashed white circles indicate the intensity
    nodes and phase singularities in the field, and the dashed white
    square indicates the region plotted in (b). Streamlines show the
    direction of the phase gradient, along which nonconservative
    forces act.}
  \label{fig:phasePlot}
\end{figure}

Not all zero-momentum waves are standing waves.  Some have non-trivial
phase profiles and so can exert nonconservative forces.  The archetype
for such pseudo-standing waves is a superposition of three plane waves
with equal amplitude, $p_0/3$, and wave vectors
\begin{equation}
  \vec{k}_n =
  -k \left[
    \cos\left(n \frac{2 \pi}{3} \right) \, \hat{x}
    +
    \sin\left(n \frac{2 \pi}{3} \right) \, \hat{y}
  \right]
\end{equation}
that satisfy $\sum_{n=1}^3 \vec{k}_n = 0$ \cite{masajada2001optical}.
The pressure intensity for such a three-wave superposition is plotted
in Fig.~\ref{fig:phasePlot}(a) and displays a six-fold rotational
symmetry similar to that of the corresponding six-fold standing wave
in Fig.~\ref{fig:intplots}(b).  The triangular lattice of antinodes in
Fig.~\ref{fig:phasePlot}(a), however, is meshed with a dual hexagonal
lattice of nodes, as indicated by dotted circles in
Fig.~\ref{fig:phasePlot}(a).  Because nonconservative forces tend to
be weaker than conservative forces by a factor of $(k a_p)^3$, we
focus our attention on regions where the sound field forms stable
traps, and expand about these points in polar coordinates,
$\vec{r} = (r, \theta)$ for small displacements, $kr < 1$.

For the three-fold pseudo-standing wave, the conservative
contributions from Eqs.~\eqref{eq:falpha} and
\eqref{eq:fbetaconservative} simplify to
\begin{align}
  \label{eq:pseudostandingconservative}
  \vec{F}^c(\vec{r}) 
  = 
  \frac{1}{8} 
  \left(2\alpha_a' -  \beta_a' \right) \nabla u^2.
\end{align}
Particles satisfying
\begin{equation}
  \label{pseudoantinode}
  2 \alpha_a' - \beta_a' > 0
\end{equation}
therefore are drawn to pressure antinodes while complementary
particles seek out nodes.

A node-seeking particle experiences amplitude and phase profiles of
the form
\begin{subequations}
  \begin{align}
    u_\text{node}(\vec{r})
    & \approx
      \frac{1}{2} p_0 \, kr 
      \quad \text{and} \\
    \varphi_\text{node}(\vec{r})
    & \approx
      \pm\theta - \frac{\pi}{2}.
      \label{eq:nodephase}
  \end{align}
  The conservative part of the associated force field exerts a Hookean
  restoring force,
  \begin{equation}
    \label{eq:nodeconservative}
    \vec{F}_\text{node}^c(\vec{r})
    \approx
    -\frac{1}{16} k p_0^2
    (\beta_a' - 2\alpha_a')\, k r \, \hat{r},
  \end{equation}
  that keeps the particle localized near the node.
  The force field also includes a nonconservative component,
  \begin{align}
    \vec{F}^{nc}_\text{node}(\vec{r})
    \approx 
    \pm\frac{1}{8} k p_0^2 (\alpha_a''+ \beta_a'') \, kr \, 
    \hat{\theta} ,
  \end{align}
\end{subequations}
that is directed by the azimuthal phase gradient and causes the
displaced particle to orbit its node.  The pseudo-standing wave
therefore transfers orbital angular momentum to particles moving near
its nodes, with each node acting as a unit-charge acoustic vortex
\cite{zhang2011angular}.  The sign of the orbital angular momentum
alternates from site to site on the honeycomb lattice of nodes.  The
array of alternating acoustic vortexes in a pseudo-standing wave
therefore carries no net angular momentum \cite{bliokh2019spin}.

Nodes repel particles satisfying $2 \alpha_a' > \beta_a'$, which
instead seek out the triangular lattice of antinodes.  Near an
antinode, the sound field's amplitude and phase profiles are
\begin{subequations}
  \begin{align}
    u_\text{antinode}(\vec{r}) 
    & \approx 
      p_0 \left[1 - \frac{1}{4} (kr)^2
      \right] 
      \quad \text{and} \\
    \varphi_\text{antinode}(\vec{r}) 
    & \approx
      \frac{1}{24} (kr)^3 \, \cos 3 \theta .
  \end{align}
  For small displacements, the conservative terms from
  $\vec{F}_\alpha(\vec{r})$ and $\vec{F}_\beta(\vec{r})$ exert a
  Hookean restoring force on an antinode-seeking particle:
  \begin{equation}
    \label{eq:antinodeconservative}
    \vec{F}_\text{antinode}^c(\vec{r}) 
    = 
    -\frac{1}{8} k p_0^2
    (2\alpha_a' - \beta_a') \, kr \, \hat{r}.
  \end{equation}
  The nonconservative terms create a sextupole flow that tends to
  drive the particle from one antinode to another:
  \begin{equation}
    \label{eq:antinodenonconservative}
    \vec{F}_\text{antinode}^{nc}(\vec{r})
    =
    F_0 \, k^2 r^2 
    (\cos 3 \theta \, \hat{r} -
    \sin 3 \theta \, \hat{\theta}),
  \end{equation}
\end{subequations}
where $F_0 = k p_0^2 \, (2 \alpha_a'' - \beta_a'') / 32$.

The conservative part of the pseudo-standing wave's force field
vanishes for materials satisfying $2\alpha_a' = \beta_a'$, leaving a
purely nonconservative force field.  This contrasts with the force
exerted by a standing wave, which is always conservative.
Realizing this condition in practice, however, would require a precise
balance of material properties.

More generally, systems satisfying
$2\alpha_a''-\beta''_a > 2 \alpha_a' - \beta_a'$ will experience
nonconservative forces that rival conservative trapping forces.  In
the particular case of an air bubble of size $k a_p = \num{0.3}$ in
water, for example, the nonconservative force exceeds the conservative
force for displacements greater than $kr \approx \num{0.025}$ from the
pressure antinodes.  A pseudo-standing wave at amplitude
$p_0 = \SI{1}{\kilo\pascal}$ and frequency
$f = \omega/(2 \pi) = \SI{2}{\mega\hertz}$ yields a nonconservative
force on the order of \SI{1}{\nano\newton} for a displacement of
$kr = \num{0.1}$ and a conservative force of just
\SI{0.1}{\nano\newton}.  The overall force field, being mostly
nonconservative, resembles the streamlines in
Fig.~\ref{fig:phasePlot}.

These observations illustrate that nonconservative forces can emerge
along directions where the sound field carries no net momentum.  More
generally, it shows that radiation pressure can be directed
independently of the direction of wave propagation.  This independence
can be used to craft tractor beams from propagation-invariant Bessel
beams \cite{marston06,lee10,yevick16}.

\subsection{Bessel Beams and Tractor Beams}
\label{sec:bessel}

Both the standing wave and the pseudo-standing wave require boundary
conditions that completely enclose their targets.  Long-range
manipulation without physical confinement is best achieved with
propagation-invariant traveling waves.  The natural basis for such
applications is the family of Bessel beams
\cite{durnin87,durnin87a,marston06}, which are non-diffracting
solutions to Eq.~\eqref{eq:waveequation} in cylindrical coordinates,
$\vec{r} = (r, \theta, z)$.  The amplitude and phase profiles for a
Bessel beam propagating along $\hat{z}$ are
\begin{align}
  u_{\gamma,n}(\vec{r})
  & = 
    p_0 \, J_n(k r \sin \gamma) 
    \quad \text{and} \\
  \varphi_{\gamma,n}(\vec r)  
  & = 
    kz \cos\gamma + n\theta ,
\end{align}
respectively, where $J_n(\cdot)$ is a Bessel function of the first
kind of order $n$.  Bessel beams are distinguished by the convergence
angle $\gamma$ that ranges from $\gamma = 0$ for conventional plane
waves to $\gamma = \pi/2$ for circular standing waves, and the integer
$n$ that imposes a helical pitch on the beam's wavefronts.

The conservative part of a Bessel beam's force field is directed
radially.  Beams with $n = 0$ have maximum intensity along the axis,
$r = 0$.  Those with $n > 0$ have zero intensity on the axis.  The
radial component of the acoustic force is linear in $kr$ for
$\abs{n} \leq 2$ and scales as $(kr)^3$, or higher, for $\abs{n} > 2$.
Whether the force attracts the particle to the axis or repels it
depends on the particle's properties.  For simplicity, we restrict our
analysis to $\abs{n} \leq 2$, in which case the conditions for
particles to be trapped on-axis are
\begin{subequations}
  \begin{align}
    \label{eq:n0trapping}
    4\alpha_a' + \beta_a' \, (1 + 3\cos 2\gamma) 
    & > 0, 
    & n & = 0 \\
    \alpha_a' + \beta_a'\, \cos 2\gamma 
    & < 0, 
    & n & = \pm 1 \\
    \beta_a' & < 0, 
    & n & = \pm 2.
  \end{align}
\end{subequations}
As noted recently \cite{fan2019trapping}, the condition for stable
trapping by an $n = 0$ Bessel beam differs qualitatively from the
analogous condition in Eq.~\eqref{eq:standingantinode} for trapping at
an antinode of a standing wave.  For example, dense objects with large
values of $\beta_a'$ are repelled by the antinodes of standing waves
and pseudo-standing waves, but tend to be trapped by the central
antinode of a Bessel beam with $\gamma < \pi/4$.  Similar reversals
arise for trapping at the central node of Bessel beams with
$\abs{n} > 0$.

Having established the condition for stable transverse trapping, we
next analyze the axial force on a particle localized at $r = 0$.  Any
beam satisfying $\vec{F}(\vec{r})\vert_{r = 0} \cdot \hat{z} < 0$ can
be said to act as a tractor beam.  It should be noted that optical
Bessel beams do not act as tractor beams for small objects because the
dipole-order photokinetic force is always repulsive \cite{yevick16}.

\begin{figure}
  \centering \includegraphics[width=\columnwidth]{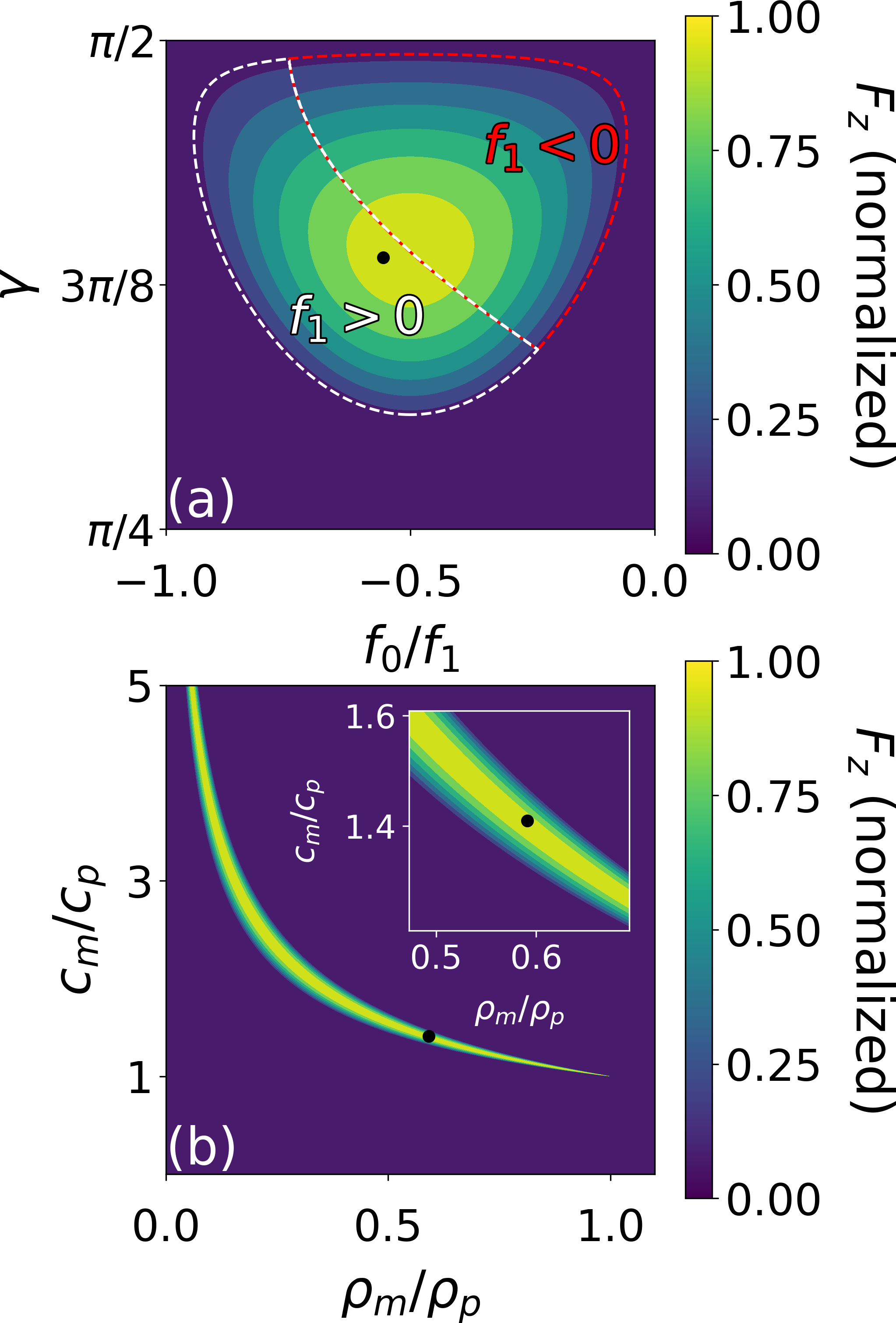}
  \caption{Relative strength of the axial force,
    $F_z \equiv \vec{F}(\vec{r})\vert_{r = 0} \cdot \hat{z}$, acting
    on an object that is trapped along the axis of an $n = 0$ Bessel
    beam. A value of 0 indicates that the object is not trapped
    on-axis.  (a) Dense objects with $f_1 > 0$ are pulled upstream by
    the Bessel beam under conditions enclosed by the white curve and
    are repelled under complementary conditions enclosed by the red
    curve. These conditions are reversed for buoyant objects with
    $f_1 < 0$.  Density-matched objects are always repelled.  The
    Bessel beam acts like a tractor beam for all other objects along
    the interface between the red- and white-bordered regions.  (b)
    The axial pulling force as a function of material parameters for
    the specific case $\gamma = \ang{70}$.  The black point indicates
    parameters for droplets of the industrial solvent xylene
    hexafluoride in butanol.  Inset: the region near this system.  }
  \label{fig:trapPull}
\end{figure}

Axial forces in propagation-invariant Bessel beams are inherently
nonconservative.  The relevant terms in Eq.~\eqref{eq:acoustokinetic}
yield
\begin{subequations}
  \begin{align}
    \label{eq:Bessel00}
    \vec{F}(\vec{r})\vert_{\substack{n=0\\r=0}} 
    & =
      \frac{1}{2} \left[
      (\alpha_a''+\beta_a'') \, u^2 + 
      \frac{\beta_a''}{2k^2} \nabla^2 u^2  \right] \, 
      \nabla\varphi \\
    & =
      \frac{1}{2} \, p_0^2 k \, (\alpha_a'' + 
      \beta_a'' \, \cos^2\gamma) \, \cos\gamma \, 
      \hat{z} .
  \end{align}
\end{subequations}
Unlike optical Bessel beams, therefore, acoustic Bessel beams can act
as tractor beams for particles satisfying both the trapping condition
from Eq.~\eqref{eq:n0trapping} and also
\begin{equation}
  \label{eq:n0tractor}
  \alpha_a'' + \beta_a'' \, \cos^2\gamma < 0.
\end{equation}
Expressed in terms of material properties, these conditions simplify
to
\begin{subequations}
  \label{eq:reversalCond}
  \begin{gather}
    f_0 < \frac{3}{8} (1 + 3 \cos 2\gamma )f_1
    \quad \text{and} \\
    \left(\frac{f_0}{f_1}\right)^2 + \frac{f_0}{f_1} + \frac{3}{4}
    \cos^2\gamma < 0.
  \end{gather}
\end{subequations}
Figure~\ref{fig:trapPull} shows the domain of beam shapes and particle
compositions for which the $n = 0$ Bessel beam acts as a tractor beam.
These include the optimal condition $f_0/f_1=-1/2$ that was identified
in the original discussion of acoustic tractor beams \cite{marston06}.

Density-matched objects ($f_1 = 0$) can only be trapped if they are
compressible enough that $f_0 < 0$.  This means, however, that
$\vec{F}\cdot \hat{z} > 0$, from which we conclude that Bessel beams
are not tractor beams for such objects.  This is reflected in
Fig.~\ref{fig:trapPull}(b), which presents the axial pulling force as
a function of the relative sound speed, $c_m/c_p$, and density,
$\rho_m/\rho_p$ for a strongly converging Bessel beam with
$\gamma = \ang{70}$.

For this convergence angle, the previously discussed system of xylene
hexafluoride droplets immersed in butanol will experience a tractor
force in the Rayleigh regime.

The analogous treatment for $n = 1$ yields
\begin{subequations}
  \begin{align}
    \vec{F}(\vec{r})\vert_{\substack{n=1\\r=0}}
    & =  
      \frac{\beta_a''}{4k^2} \, 
      \nabla^2 u^2 \, \nabla \varphi\\
    & = 
      \frac{1}{4} p_0^2 k \, \beta_a'' \,
      \cos \gamma \sin^2 \gamma \, \hat{z}.
  \end{align}
\end{subequations}
Because $\beta_a'' > 0$, such beams do act as tractor beams for any
choice of materials, at least not for objects trapped on the axis.
They instead drive trapped objects downstream.

The $n=2$ beam can trap small objects on the axis, but exerts no axial
force at all,
\begin{equation}
  \vec{F}(\vec{r}) \vert_{\substack{n=2\\r=0}}
  = 0.
\end{equation}
Such beams might serve as useful conduits for Rayleigh particles that
are moved back and forth along the axis by other forces.

\section{Discussion}
\label{sec:discussion}

The theory of acoustokinetic forces presented here expresses the
influence of a sound wave on a small object in terms of the amplitude
and phase profiles of the pressure field.  The resulting analytical
framework, which is summarized in Eq.~\eqref{eq:acoustokinetic},
offers a complementary perspective on acoustic forces to the standard
development, which explicitly refers to the velocity field.
Acoustokinetic expressions are particularly useful for designing
acoustic force fields because they inherently account for coupling
between the pressure and velocity fields and specify amplitude and
phase profiles that can be controlled with transmissive, reflective or
emissive elements.

The acoustokinetic approach yields expressions that closely resemble
results for optical forces in the Rayleigh limit.  Exploring the
differences and similarities between acoustokinetic and photokinetic
forces offers insights into how sound fields couple to material
properties to generate useful and interesting force landscapes.  The
most important difference is that the particle's polarizability
includes contributions from the incident wave's monopole component.
This means that the force arising from the sound field's dipole
component includes terms at the same order of magnitude as leading
quadrupole contributions.  Such mixing does not arise in optical
forces.  It vanishes for objects that are density matched with the
medium, yielding expressions for acoustic forces that are exactly
analogous to their optical counterparts.

The acoustokinetic framework naturally accounts for the conservative
nature of the force field exerted by standing waves and differentiates
it from the influence of pseudo-standing waves that also carry no net
momentum yet still exert nonconservative forces.  When applied to
acoustic Bessel beams, the acoustokinetic framework provides clear
guidance on the limits for axial trapping and the conditions under
which a zeroth-order Bessel beam can act as a tractor beam.  This
result contrasts with the equivalent analysis of optical Bessel beams,
which do not act as tractor beams in the Rayleigh regime.

These examples illustrate the value of the acoustokinetic framework
for analyzing acoustic force fields.  More generally, the expressions
from Eq.~\eqref{eq:acoustokinetic} can be solved either numerically,
or in some cases analytically, for the phase and amplitude profiles
that correspond to a specified force field acting on a particle of a
given size, compressibility and density.  As in the optical case, the
framework can be extended beyond the dipole approximation.  For
objects substantially smaller than the wavelength of sound, however,
the dipole-order terms are simple enough to yield analytical
expressions.  This framework also can be extended to handle
interactions among insonated objects, an effect usually referred to as
secondary Bjerknes forces.

\section{Acknowledgments}
\label{sec:acknowledgments}

This work was supported by the MRSEC program of the National Science
Foundation through Award Number DMR-1420073.  The authors acknowledge
helpful conversations with Phillip Marston (Washington State
University) and Konstantin Bliokh (The Australian National University
and RIKEN, Japan).

\appendix
\section{Derivation of Equation~\eqref{eq:acoustokinetic}}
\label{sec:appendix}

The central idea of the acoustokinetic framework is to factor the
incident pressure field $p(\vec r)$ into its real-valued amplitude
$u(\vec r)$ and phase $\varphi(\vec r)$:
\begin{equation}
  p(\vec{r}) = u(\vec{r}) \, e^{i \varphi(\vec{r})}.
  \label{eq:factorization}
\end{equation}
Expressed in these terms, the Helmholtz wave equation,
Eq.~\eqref{eq:helmholtz}, separates into two coupled differential
equations that constrain the amplitude and phase profiles,
\begin{subequations}
  \label{eq:helm12}
  \begin{gather}
    \nabla^2 u + k^2 u - u(\nabla \varphi)^2 = 0 \label{eq:helm1} \\
    u \nabla^2 \varphi + 2 \nabla u \cdot \nabla \varphi =
    0. \label{eq:helm2}
  \end{gather}
\end{subequations}
These constraints transform the expression for the time-averaged
acoustic force in Eq.~\eqref{eq:silva} into the acoustokinetic
expressions in Eq.~\eqref{eq:acoustokinetic}.

The right-hand side of Eq.~\eqref{eq:silva} is readily expanded into
the sum of four terms
\begin{subequations}
  \label{eq:silvaexpanded}
  \begin{align}
    \vec{F}(\vec{r}) = 
    \frac{1}{2} \big[
    & \quad 
      \alpha_a' \, \Re\{p\nabla p^*\} - \alpha_a'' \, \Im\{p\nabla p^*\} 
      \label{eq:forcea} \\
    & + k^{-2}\beta_a' \, \Re\{(\nabla p \cdot \nabla) \nabla p^*\} 
      \label{eq:forceb} \\
    & - k^{-2}\beta_a'' \, \Im\{(\nabla p \cdot \nabla) \nabla p^*\}\big],
      \label{eq:forcec}
  \end{align}
\end{subequations}
each of will be expressed in terms of $u(\vec{r})$ and
$\varphi(\vec{r})$.  Substituting Eq.~\eqref{eq:factorization} into
Eq.~\eqref{eq:forcea} and noting that
\begin{equation}
  p\nabla p^* 
  =
  \frac{1}{2}\nabla u^2 - i u^2 \nabla \varphi
  \label{eq:pnablap}  
\end{equation}
leads directly to Eq.~\eqref{eq:falpha}.  The first term on the
right-hand side of Eq.~\eqref{eq:pnablap} is the gradient of an
analytic function, and therefore corresponds to a manifestly
conservative force.  We demonstrate that the second term corresponds
to a purely nonconservative force by showing that it is
divergence-free:
\begin{equation}
  \begin{aligned}
    \nabla \cdot (u^2 \nabla \varphi)
    & = \nabla(u^2) \cdot \nabla \varphi + u^2 \nabla^2 \varphi \\
    & = 2 u \nabla u \cdot \nabla \varphi + u^2 \nabla^2 \varphi \\
    & = 0,
  \end{aligned}
\end{equation}
with the last line following from Eq.~\eqref{eq:helm2}.

The expressions in Eqs.~\eqref{eq:forceb} and \eqref{eq:forcec} depend
on the real and imaginary parts of
$(\nabla p \cdot \nabla)\nabla p^*$.  The real part is conveniently
transformed using the vector identity,
\begin{multline}
  \nabla (\vec{A} \cdot \vec{B}) = (\vec{A} \cdot \nabla) \vec{B}
  + (\vec{B} \cdot \nabla) \vec{A} \\
  + \vec{A} \times (\nabla \times \vec{B}) + \vec{B} \times (\nabla
  \times \vec{A}).
  \label{eq:identity1}
\end{multline}
Setting $\vec{A} = \nabla p$ and $\vec{B} = \nabla p^*$ causes the
curls on the right-hand side of Eq.~\eqref{eq:identity1} to vanish
identically.  The two remaining terms yield
\begin{align}
  \Re\{(\nabla p \cdot \nabla)\nabla p^*\} 
  & = 
    \frac{1}{2} \nabla(\nabla p \cdot \nabla p^*) \\
  & =
    \frac{1}{2} \nabla\left[ (\nabla u)^2 
    + u^2 (\nabla \varphi)^2 \right].
\end{align}
Combining this with Eq.~\eqref{eq:helm1} and the identity
$\nabla^2(u^2)=2u\nabla^2u+2(\nabla u)^2$ yields
Eq.~\eqref{eq:fbetaconservative}.

Equation~\eqref{eq:fbetanonconservative} similarly follows from the
identity
\begin{multline}
  \nabla \times (\vec{A} \times \vec{B}) =
  \vec{A} (\nabla \cdot \vec{B}) - \vec{B} (\nabla \cdot \vec{A}) \\
  + (\vec{B} \cdot \nabla) \vec{A} -(\vec{A} \cdot \nabla) \vec{B} .
\end{multline}
Setting $\vec{A} = \nabla p$ and $\vec{B} = \nabla p^*$, we obtain
\begin{multline}
  \Im\{(\nabla p \cdot \nabla)\nabla p^*\} =
  - \Im\{ \nabla^2p\nabla p^* \} \\
  + \frac{i}{2}\nabla \times (\nabla p \times \nabla p^*) .
  \label{eq:lastpart}
\end{multline}
We use the Helmholtz equation to eliminate the Laplacian operator on
the right-hand side so that
\begin{align}
  \Im\{\nabla^2 p \nabla p^*\} 
  & = k^2 \Im\{p \nabla p^\ast\} \\
  & = - k^2 u^2 \nabla \varphi,
    \label{eq:lastline}
\end{align}
with Eq.~\eqref{eq:lastline} following from Eq.~\eqref{eq:pnablap}.
As before, this term is divergence-free and therefore corresponds to a
purely non-conservative force.

The second term on the right-hand side of Eq.~\eqref{eq:lastpart} is
the curl of a function and thus also corresponds to a non-conservative
force.  It may be expressed in terms of the wave's amplitude and phase
profiles as
\begin{multline}
  \frac{i}{2}\nabla \times (\nabla p \times \nabla p^*) = \nabla
  \times
  \left[ u(\nabla u \times \nabla \varphi)\right] \\
  = u \nabla \times (\nabla u \times \nabla \varphi) + \nabla u \times
  (\nabla u \times \nabla \varphi)
  \label{eq:tripleproduct}
\end{multline}
The first term on the right-hand side of Eq.~\eqref{eq:tripleproduct}
can be rewritten as
\begin{multline}
  \label{eq:curlterm}
  u \nabla \times (\nabla u \times \nabla \varphi) =
  u (\nabla^2 \varphi + \nabla \varphi \cdot \nabla)\nabla u \\
  -u (\nabla^2 u + \nabla u \cdot \nabla) \nabla \varphi,
\end{multline}
while the second can be expressed as
\begin{multline}
  \nabla u \times (\nabla u \times \nabla \varphi) =
  (\nabla u \cdot \nabla \varphi) \nabla u \\
  - (\nabla u \cdot \nabla u) \nabla \varphi.
\end{multline}
Using Eq.~\eqref{eq:helm2} to eliminate
$\nabla u \cdot \nabla \varphi$ and combining all remaining terms
yields Eq.\eqref{eq:fbetanonconservative}.

%

\end{document}